\pdfoutput=1
\documentclass[journal,a4paper]{IEEEtran}
\usepackage[utf8]{inputenc}

\usepackage{graphicx}
\DeclareGraphicsExtensions{.png,.pdf}
\graphicspath{{img/}}
\usepackage[caption=false,font=footnotesize]{subfig}
\usepackage{dblfloatfix}

\usepackage{cite}

\usepackage{tikz}
\usetikzlibrary{positioning} 
\usetikzlibrary{fit} 
\usetikzlibrary{calc} 
\usetikzlibrary{shapes} 
\usetikzlibrary{backgrounds} 
\usepackage{pgfplots}
\pgfplotsset{compat=1.12}
\usepgfplotslibrary{colorbrewer}

\usepackage{cite}

\usepackage{balance}
\usepackage[cmex10]{amsmath}
\usepackage{amssymb}
\usepackage{amsbsy}

\newcounter{lemma}

\newcommand{\vect}[1]{\ensuremath{\mathit{\boldsymbol{#1}}}}

\usepackage{acronym}

\acrodef{ttcm}[TTCM]{turbo trellis-coded modulation}
\acrodef{ldpc}[LDPC]{low density parity check}
\acrodef{cm}[CM]{coded codulation}
\acrodef{bicm}[BICM]{bit-interleaved coded modulation}
\acrodef{bcjr}[BCJR]{Bahl, Cocke, Jelinek, and Raviv}
\acrodef{rsc}[RSC]{recursive systematic convolutional}
\acrodef{brgc}[BRGC]{binary reflected Gray code}

\acrodef{nb}[NB]{nonbinary}
\acrodef{fec}[FEC]{forward error correction}
\acrodef{nbfec}[NB-FEC]{\ac{nb} \ac{fec}}
\acrodef{hdfec}[HD-FEC]{hard decision \ac{fec}}
\acrodef{sdfec}[SD-FEC]{soft decision \ac{fec}}
\acrodef{oh}[OH]{overhead}
\acrodef{dsp}[DSP]{digital signal processing}
\acrodef{hd}[HD]{hard decision}
\acrodef{sd}[SD]{soft decision}

\acrodef{mi}[MI]{mutual information}
\acrodef{gmi}[GMI]{generalized mutual information}
\acrodef{air}[AIR]{achievable information rate}

\acrodef{awgn}[AWGN]{additive white Gaussian noise}
\acrodef{psk}[PSK]{phase-shift keyed}

\acrodef{snr}[SNR]{signal to noise ratio}
\acrodef{osnr}[OSNR]{optical signal to noise ratio}
\acrodef{se}[SE]{spectral efficiency}
\acrodef{ber}[BER]{bit error rate}

\acrodef{ecl}[ECL]{external cavity laser}
\acrodef{cw}[CW]{continues wave}
\acrodef{mzm}[MZM]{Mach--Zehnder modulator}
\acrodef{dac}[DAC]{digital analog converter}
\acrodef{dp}[DP]{dual polarisation}
\acrodef{pbs}[PBS]{polarisation beam splitter}
\acrodef{ssmf}[SSMF]{standard single mode fibre}
\acrodef{aom}[AOM]{acousto-optic modulator}

\acrodef{qam}[QAM]{quadrature amplitude modulation}
\acrodef{qpsk}[QPSK]{quadrature phase shift keying}
\acrodef{8psk}[8PSK]{8-ary phase-shift keying}

\acrodef{edfa}[EDFA]{Erbium doped fibre amplifier}
\acrodef{ase}[ASE]{amplified spontaneous emission}

\acrodef{iid}[i.i.d.]{independent and identically distributed}
\acrodef{cs}[CS]{circular symetric}
\acrodef{ncs}[NCS]{non circular symetric}
\acrodef{pdf}[PDF]{probability density function}

\acrodef{ml}[ML]{maximum likelihood}
\acrodef{mll}[MLL]{minus log likelihood}
\acrodef{ll}[LL]{log likelihood}
\acrodef{llr}[LLR]{log likelihood ratio}

\acrodef{bch}[BCH]{Bose Ray-Chaudhuri Hocquenghem}
\acrodef{rs}[RS]{Reed Solomon}
\acrodef{cc}[CC]{convolutional code}
\acrodef{tcm}[TCM]{trellis-coded modulation}

\tikzset{%
	pic shift/.store in=\shiftcoord,
    pic shift={(0,0)},
    pic fill/.store in=\fillcolor,
    pic fill=white,
    EDFA/.pic={
        \begin{scope}[shift={\shiftcoord}]
        \draw[fill=\fillcolor] (0,0) coordinate (-in) --  (0,9pt) -- (18pt,0) coordinate (-out) --  (0,-9pt) -- cycle;
        \node[anchor=north,inner sep=2pt] at (9pt,-9pt) (-label) {EDFA};
        \end{scope}
    },
    Coupler/.pic={
        \begin{scope}[ shift={\shiftcoord}]
        \draw[-] (0,0) coordinate (-in1) to[out=0,in=180] (24pt,-18pt) to[out=0,in=180] (48pt,0) coordinate (-out1);
        \draw[-] (0,-36pt) coordinate (-in2) to[out=0,in=180] (24pt,-18pt) to[out=0,in=180] (48pt,-36pt) coordinate (-out2);
        \end{scope}
    },
    Modulator/.pic={
      \begin{scope}[shift={\shiftcoord}]
	\node[anchor=center,fill=\fillcolor] (txt) at (18pt,0) {MZM};
    \draw[-] (0,0) coordinate (-in) to (3pt,6pt) to (33pt,6pt) to (36pt,0) coordinate (-out);
    	\draw[-] (-in) to (3pt,-6pt) to (33pt,-6pt) to (-out);
     	\node[draw] (dac) at (18pt,18pt) {DAC};
 		\draw[->] (dac) -- (18pt,6pt);
      \end{scope}
    },
    Attenuator/.pic={
    	\begin{scope}[shift={\shiftcoord}]
			\coordinate (-in) at  (0,0);
            \node[draw,anchor=west,minimum width=18pt, minimum height=18pt,fill=\fillcolor] at (-in) (-box) {};
            \draw  (9pt,0) circle (4pt);
            \draw[->] (3pt,-6pt) -- (15pt,6pt);
            \coordinate (-out) at (18pt,0);
		\end{scope}
    },
        Filter/.pic={
    	\begin{scope}[shift={\shiftcoord}]
			\coordinate (-in) at  (0,0);
            \node[draw,anchor=west,minimum width=18pt, minimum height=18pt,fill=\fillcolor] at (-in) (-box) {};
            \draw[-] (3pt,3pt) to[bend left] (9pt,3pt) to[bend right] (15pt,3pt);
            \draw[-] (3pt,0) to[bend left] (9pt,0) to[bend right] (15pt,0);
            \draw[-] (3pt,-3pt) to[bend left] (9pt,-3pt) to[bend right] (15pt,-3pt);
            \draw[->] (3pt,-6pt) -- (15pt,6pt);
            \coordinate (-out) at (18pt,0);
		\end{scope}
    },
      Loop/.pic={
    	\begin{scope}[shift={\shiftcoord}]
			\coordinate (-in) at  (0,0);
            \draw[->] (-in) -- +(12pt,0) node[draw, anchor=west, minimum height=18pt,fill=Set3-8-6] (aom1) {AOM 1};
            
            \node[draw, anchor=west, minimum height=18pt,fill=Set3-8-6] (aom2) at ($(aom1.west)+(0,-36pt)$) {AOM 2};
            \draw (aom1.east) pic (coupler) {Coupler};
            \draw[-] ($(aom1.east)+(48pt,0)$) -- ($(aom1.east)+(90pt,0)$) coordinate (-out);
            
            \draw[-] ($(aom1.east)+(48pt,-36pt)$) -- ($(aom1.east)+(72pt,-36pt)$);
            \draw[->] ($(aom1.east)+(72pt,-36pt)$) [rounded corners=6pt]-| ++ (18pt,-12pt) [rounded corners=6pt] |- ++ (-72pt,-24pt) coordinate (-edfa) pic[xscale=-1, rounded corners=0pt, pic fill=Set3-8-5] (edfa) {EDFA};
            \draw[->] ($(-edfa)+(-18pt,0)$) -- ++(-12pt,0) coordinate (-filer) pic[pic shift={(-18pt,0)},pic fill=Set3-8-7] (filter) {Filter};
            \draw[->] ($(-filer)+(-18pt,0)$) [rounded corners=6pt] -| (0pt,-50pt) |- (aom2.west) ;
            \draw ($(-edfa)+(24pt,0)$) to[out=0, in=135] + (6pt,-3pt);
            \draw ($(-edfa)+(24pt,0)$) to[out=180, in=90] ++ (-9pt,-12pt) coordinate (-raman);
            \node[anchor=north west, inner sep=0] (-ramanlabel) at (-raman) {Raman};
            
            \coordinate (-spools) at ($(-edfa)+(48pt,0)$);
            
            \node[draw,circle, anchor=south, minimum height=12pt] at (-spools) {};
            \node[draw,circle, anchor=south, minimum height=12pt] at ($(-spools)+(3pt,0)$) {};
            \node[draw,circle, anchor=south, minimum height=12pt] at ($(-spools)+(-3pt,0)$) {};
            \node[anchor=south east] at ($(-spools)+(18pt,12pt)$) {$75$~km SSMF};
            
            \node[draw, dashed, fit=(-in)(-out)(aom1)(-ramanlabel)] (loopbox) {};
            \node[above=0 of loopbox.north] {Recirculating loop};
          \end{scope}
    },
}

\tikzset{%
	every node/.style={font=\scriptsize},
}

\newcommand\matinv[1]{\ensuremath{#1^{\scriptscriptstyle-\!1}}}

\begin{document}

\title{An Experimental Comparison of Coded Modulation Strategies for 100~Gbit/s Transceivers \pgfplotsversion}

\author{Eric~Sillekens,~\IEEEmembership{Student Member,~IEEE},
 Alex~Alvarado,~\IEEEmembership{Senior Member,~IEEE},
 Chigo~M.~Okonkwo,~\IEEEmembership{Member,~IEEE},
 Benn~C.~Thomsen,~\IEEEmembership{Member,~IEEE}%
\thanks{E. Sillekens, A. Alvarado and B. C. Thomsen are with the Optical Networks Group, Department of Electronics and Electrical Engineering, University College London, London, WC1E 7JE, U.K.}%
\thanks{C. M. Okonkwo is with the COBRA Research Institute, Eindhoven University of Technology, Eindhoven, The Netherlands.}}%

\markboth{Preprint, \today.}{}

\maketitle

\begin{abstract}
\boldmath
Coded modulation is a key technique to increase the spectral efficiency of coherent optical communication systems.
Two popular strategies for coded modulation are turbo trellis-coded modulation (TTCM) and bit-interleaved coded modulation (BICM) based on low-density parity-check (LDPC) codes.
Although BICM LDPC is suboptimal, its simplicity makes it very popular in practice.
In this work, we compare the performance of TTCM and BICM LDPC using information-theoretic measures.
Our information-theoretic results show that for the same overhead and modulation format only a very small penalty (less than $0.1$~dB) is to be expected when an ideal BICM LDPC scheme is used.
However, the results obtained for the coded modulation schemes implemented in this paper show that the TTCM outperforms BICM LDPC by a larger margin. For a $1000$~km transmission at $100$~Gbit/s, the observed gain was $0.4$~dB.
\end{abstract}


\section{Introduction}

A promising alternative to increase the \ac{se} of optical transmission systems is to use higher order modulation formats. 
To maintain reliable communication, the decreased sensitivity caused by high order modulation formats is compensated by \ac{fec}.
The combination of a \ac{nb} modulation format and \ac{fec} is known as \ac{cm} \cite{Ungerboeck1982ChannelSignals}.
Most current 100G transceivers use \ac{qpsk} but future 400G transceivers are expected to employ \ac{cm} based on 16-\ac{qam}\cite{Xia2014TransmissionField,Huang2014TransmissionKm}.
Using higher order modulation formats is also a topic of current research, both in point-to-point links \cite{Millar2015DetectionReceiver,Maher2016IncreasingPerformance.,Buchali2014ImplementationTransmission} and in the context of optical networks \cite{LuFLEXIBLEJapan,Alvarado2015OnNetworks,Alvarado2016OnNetworks,Mello2014OpticalTransceivers}.

\ac{cm} can be implemented in several ways. The most typical approach is to separate the coding (decoding) from the mapping (demapping) functions at the transmitter (receiver). This separation has the advantage that the binary \ac{fec} can be designed independently of the modulation format. This structure is typically known as \ac{bicm} \cite{Zehavi19928-PSKChannel,FabregasBit-InterleavedBooks,Szczecinski2015Bit-InterleavedDesign}.
Another approach to \ac{cm} is to combine the \ac{fec} and mapping into a single operation at the transmitter and to pass the channel outputs directly to a \ac{nbfec} decoder to recover the data bits at the receiver. This idea dates back to Ungerboeck's celebrated \ac{tcm} \cite{Ungerboeck1982ChannelSignals}. 

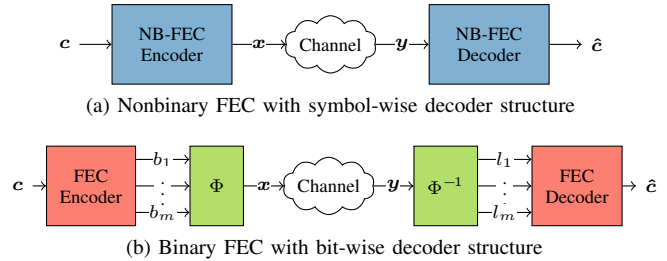
\begin{figure}[!t]
\centering

\subfloat[][Nonbinary FEC with symbol-wise decoder structure]{
 \begin{tikzpicture}[node distance=12pt,op/.style={draw,minimum width=3em, minimum height=3em,inner xsep=1em},bitlabel/.style={midway,inner sep=0,fill=white,font=\scriptsize}]

\node[draw,cloud puffs=10,cloud puff arc=120, aspect=2,cloud, inner sep=0,fill=white] (channel) {Channel};

\node[op,left=2em of channel,fill=Set3-8-5,align=center] (encoder) {NB-FEC\\Encoder};
\node[left= of encoder] (input) {$\vect{c}$};

\draw[->] (input) -- (encoder)  coordinate [midway] (inf bits);
\draw[->] (encoder) -- (channel) node [bitlabel] {$\vect{x}$};

\node[op, right=2em of channel,align=center,fill=Set3-8-5] (decoder) {NB-FEC\\Decoder};

\draw[->] (channel) -- (decoder) node [bitlabel] {$\vect{y}$};

\node[right= of decoder] (output) {$\vect{\hat{c}}$};
\draw[->] (decoder) -- (output);

\end{tikzpicture}
\label{subfig:intro_ttcm}}

\subfloat[][Binary FEC with bit-wise decoder structure]{
 \begin{tikzpicture}[node distance=1em,op/.style={draw,minimum width=2em, minimum height=3em,inner xsep=.5em},bitlabel/.style={midway,inner sep=0,fill=white,font=\scriptsize}]

\node[draw,cloud puffs=10,cloud puff arc=120, aspect=2,cloud, inner sep=0,fill=white] (channel) {Channel};

\node[op,left=1.5em of channel,fill=Set3-8-7] (mapper) {$\Phi$};
\node[op,left= 2em of mapper,fill=Set3-8-4,align=center] (encoder) {FEC\\Encoder};
\node[left=.5em of encoder] (input) {$\vect{c}$};

\draw[->] (input) -- (encoder)  coordinate [midway] (inf bits);

\draw[->] ($(encoder.east) + (0,1em)$) -- ($(mapper.west) + (0,1em)$) node[bitlabel] {$b_1$};
\draw[->] ($(encoder.east) + (0,-1em)$) -- ($(mapper.west) + (0,-1em)$) node[bitlabel] {$b_m$};
\draw[->] (encoder) -- (mapper) node [bitlabel,inner xsep=.5ex]  (enc bits) {$\vdots$};

\draw[->] (mapper) -- (channel)  node [bitlabel] {$\vect{x}$};

\node[op, right=1.5em of channel,fill=Set3-8-7] (demapper) {$\matinv{\Phi}$};
\node[op, right= 2em of demapper,align=center,fill=Set3-8-4] (decoder) {FEC\\Decoder};

\draw[->] (channel) -- (demapper) node [bitlabel] {$\vect{y}$};

\draw[->] (demapper) -- (decoder) node [bitlabel,inner xsep=.5ex] (llr enc bits) {$\vdots$};
\draw[->] ($(demapper.east) + (0,1em)$) -- ($(decoder.west) + (0,1em)$) node[bitlabel] {$l_1$};
\draw[->] ($(demapper.east) + (0,-1em)$) -- ($(decoder.west) + (0,-1em)$) node[bitlabel] {$l_m$};

\node[right=.5em of decoder] (output) {$\vect{\hat{c}}$};
\draw[->] (decoder) -- (output);
\end{tikzpicture}
\label{subfig:intro_bicm}}

 \caption{The two CM strategies considered in this work.}
\label{fig:intro}
\end{figure}

In this work, we compare the performance of these two \ac{cm} strategies for two particular implementations, as shown in Fig.~\ref{fig:intro}.
We consider \ac{8psk} as the modulation format because it offers a higher \ac{se} than \ac{qpsk}, yet a lower implementation complexity than 16\ac{qam}. Additionally, the use of \ac{psk} modulation formats is also motivated by recent results \cite{Kojima2016InvestigationDPQPSK}, where they are shown to outperform \ac{qam} formats in  highly nonlinear channels (e.g.,  in dispersion-managed links). Furthermore, the codes rate we consider is $R=2/3$, which when combined with \ac{8psk} results in a \ac{se} comparable to  traditional \ac{qpsk}-based systems.

The first strategy is shown in Fig.~\ref{fig:intro}~\subref{subfig:intro_ttcm} and is based on a symbol-wise receiver structure.
Here, the encoder is a \ac{nbfec} that transforms data bits ($\vect{c}$) directly into nonbinary constellation symbols ($\vect{x}$).
After transmission, a \ac{nbfec} decoder uses the received symbols ($\vect{y}$) to retrieve the data bits ($\vect{\hat{c}}$).
The \ac{nbfec} encoder in Fig.~\ref{fig:intro}~\subref{subfig:intro_ttcm} operates on a symbol level.
The second strategy, shown in Fig.~\ref{fig:intro}~\subref{subfig:intro_bicm}, is a suboptimal implementation of the \ac{nbfec} decoder in Fig.~\ref{fig:intro}~\subref{subfig:intro_ttcm}. 
This strategy is based on a bit-wise receiver, also known as \ac{bicm} \cite{FabregasBit-InterleavedBooks,Szczecinski2015Bit-InterleavedDesign}.
At the transmitter, a binary \ac{fec} encoder converts data bits ($\vect{c}$) into encoded bits ($\vect{b}=[b_1, \ldots, b_m]^\mathrm{T}$), which are then mapped to constellation symbols ($\vect{x}$) using a memoryless mapper ($\Phi$).
These symbols are then transmitted over the channel.
In \ac{bicm},  the demapper ($\Phi^{-1}$) computes soft information on the encoded bits ($\vect{l}=[l_1,\ldots,l_m]^\mathrm{T}$) using the received symbols ($\vect{y}$).
This soft information is then passed to the binary \ac{fec} decoder to retrieve the data bits ($\vect{\hat{c}}$).
{The suboptimality of this strategy originates from the reduction of soft information caused by the bit-wise demapper, i.e., the loss caused by replacing $2^m$ symbol likelihoods (for every possible transmitted symbol) by $2m$ bit likelihoods (for every transmitted bit), thereby passing less information to the decoder to estimate the transmitted bits.}

For the \ac{nb}-\ac{fec}, we consider the \ac{8psk}-based \ac{ttcm} encoder from \cite{Robertson1998}, where each transmitted symbol carries two data bits.
At the receiver, we use a symbol-wise \emph{iterative} decoder that approximates the \ac{ml} decision rule.

The binary \ac{fec} in Fig.~\ref{fig:intro}~\subref{subfig:intro_bicm} can be any binary code.
In this work, a rate $R=2/3$ \ac{ldpc} code is considered.
\ac{ldpc} codes have recently received a great deal of attention due to their excellent performance \cite{Schmalen2015ASlips,Smith2010FutureCommunications}.
Furthermore, we consider an \ac{8psk} constellation based on the \ac{brgc} \cite{Gray53,Agrell2007GrayNoise}.
The encoded bits ($\vect{b}=[b_1,b_2,b_3]^\mathrm{T}$) are then mapped to \ac{8psk} symbols, giving a net data rate of 2~bit/symbol.
This is the same rate achieved by \ac{ttcm} in Fig.~\ref{fig:intro}~\subref{subfig:intro_bicm}.

Previously, \ac{ttcm} has been shown to improve the performance of direct detection systems \cite{Kiasaleh1998Turbo-codedSystems}, however, its performance was only compared to uncoded transmission. 
In our previous work \cite{Sillekens2015ExperimentalKeying}, the performance of the iterative \ac{ttcm} scheme discussed here, and shown in Fig.~\ref{fig:intro}~\subref{subfig:intro_ttcm}, was compared with uncoded \ac{qpsk} and also with noniterative \ac{tcm} with \ac{8psk}  \cite{Ungerboeck1982ChannelSignals} at the same information rate of 2~bit/symbol. The results of \cite{Sillekens2015ExperimentalKeying} showed that iterative decoding provided the largest performance gain.

In this paper we consider two schemes that employ iterative decoding, and thus, are comparable in terms of decoding complexity. We investigate the benefits of \ac{ttcm} over the more popular \ac{bicm} scheme. An experimental comparison between these two schemes at a net data rate of $100$~Gbit/s is presented for a \ac{dp} $1000$~km recirculating loop setup. The main contribution of this paper is to present this comparison based on information-theoretical metrics. In this paper we also present ready-to-use Monte Carlo expressions to evaluate these information-theoretical quantities.

This paper is organized as follows. In Sec.~\ref{sec:cm}, the implementation of the \ac{cm} coded modulation strategies is detailed. In Sec.~\ref{sec:method} the system performance metrics are explained and the description of the experimental setup is given in Sec.~\ref{sec:setup}. The results are presented in Sec.~\ref{sec:results} and the conclusion in Sec.~\ref{sec:conclusions}.

\section{Coded modulation strategies}
\label{sec:cm}

In this section, the implementation of the \ac{ttcm} and \ac{ldpc} schemes is described. The selection of the codeword length for both strategies is also discussed.

\subsection{\ac{ttcm}}
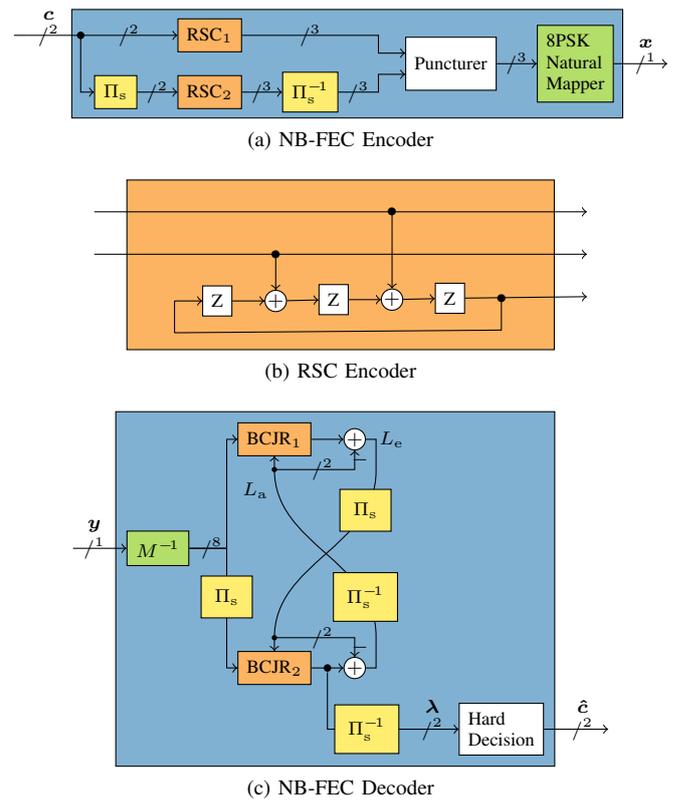
\begin{figure}[tbp]
  \centering
\subfloat[][NB-FEC Encoder]{
\begin{tikzpicture}


\coordinate (origin);

\draw[-] (origin) -- +(-5pt,0);

\coordinate[right=20pt of origin] (fork);
\draw[-] (origin) -- (fork) node[pos=0.4, label={[yshift=-.5em]above:$\vect{c}$}](c) {$/^2$};

\node[fill,circle,inner sep=1pt] at (fork) {};

\node[below right=15pt and 5pt of fork, draw,fill=Set3-12-12] (interleaver) {$\Pi_\mathrm{s}$};

\draw[->] (fork) |- (interleaver);

\node[right=15pt of interleaver,draw,fill=Set3-8-6] (rsc bot) {RSC$_2$};
\draw[->] (interleaver) -- (rsc bot) node[midway] {$/^2$};

\node[right=15ptof rsc bot,draw,fill=Set3-12-12] (deinterleaver) {$\Pi_\mathrm{s}^{\scriptscriptstyle-\!1}$};
\draw[->] (rsc bot) -- (deinterleaver) node[midway] {$/^3$};

\node[draw,fill=Set3-8-6] (rsc top) at (fork -| rsc bot) {RSC$_1$};
\draw[->] (fork) -- (rsc top) node[midway] {$/^2$};

\coordinate[right=15ptof deinterleaver] (helper1);

\coordinate (helper2) at ($(rsc top)!.5!(rsc bot)$);
\coordinate (merge) at (helper1 |- helper2);

\node[right=10pt of merge,draw,minimum height=2em,fill=white] (punc) {Puncturer};

\draw[->] (rsc top) -|  node[pos=0.25] {$/^3$} ($(merge) + (0,4pt)$) --($(punc.west) + (0,4pt)$);
\draw[->] (deinterleaver) -|  node[pos=0.25] {$/^3$} ($(merge) + (0,-4pt)$) --($(punc.west) + (0,-4pt)$);

\node[right=15pt of punc, draw, align=left,fill=Set3-8-7] (mapper) {8PSK \\ Natural \\ Mapper};
\draw[->] (punc) -- (mapper) node[midway] {$/^3$};

\coordinate[right=20pt of mapper] (symbols);

\draw[->] (mapper) -- (symbols) node[pos=.6, label={[xshift=0em,yshift=-.5em]above:$\vect{x}$}] (x){$/^1$};


 \begin{scope}[on background layer]  
\node[draw,fit=(fork)(rsc top)(interleaver)(mapper),fill=Set3-8-5] (box) {};
\end{scope}


\end{tikzpicture}
\label{subfig:ttcm_enc}}

\subfloat[][RSC Encoder]{
\begin{tikzpicture}

    \begin{scope}[scale=1, transform shape]

      \begin{scope}[on background layer]
        \node[draw,minimum height=64pt, minimum width=160pt,fill=Set3-8-6] (rsc) {};
      \end{scope}

      \coordinate (rsc b1 in) at ($(rsc.west)+(-12pt,20pt)$);
      \coordinate (rsc b2 in) at ($(rsc.west)+(-12pt, 4pt)$);

      \coordinate (rsc b1 out) at ($(rsc.east)+(12pt,20pt)$);
      \coordinate (rsc b2 out) at ($(rsc.east)+(12pt, 4pt)$);
      \coordinate (rsc b3 out) at ($(rsc.east)+(12pt,-12pt)$);

      \draw[->] (rsc b1 in) to (rsc b1 out); 
      \draw[->] (rsc b2 in) to (rsc b2 out); 

      \node[left=32pt of rsc b3 out,anchor=mid] (rsc node3) {$\bullet$};

      \node[draw, left=8pt of rsc node3,fill=white] (z3) {Z};

      \node[inner sep=0pt,draw,left=16pt of z3,anchor=mid,circle,fill=white] (rsc node2) {$+$};

      \node[draw, left=12pt of rsc node2,fill=white] (z2) {Z};

      \node[inner sep=0pt,draw,left=16pt of z2,anchor=mid,circle,fill=white] (rsc node1) {$+$};

      \node[draw, left=12pt of rsc node1,fill=white] (z1) {Z};

      \draw[->] (z3) to (rsc b3 out);

      \draw[->] (z2) to (rsc node2);
      \draw[->] (z1) to (rsc node1);

      \draw[->] (rsc node1) to (z2);
      \draw[->] (rsc node2) to (z3);

      \draw[->] (rsc node3.center) to ($(rsc node3)+(0,-12pt)$) ($(rsc node3)+(0,-12pt)$) to ($(z1)+(-16pt,-12pt)$) ($(z1)+(-16pt,-12pt)$) to ($(z1)+(-16pt,0)$) ($(z1)+(-16pt,0)$) to (z1);

      \node at (rsc b2 in -| rsc node1) (rsc branch b2) {$\bullet$}; 
      \node at (rsc b1 in -| rsc node2) (rsc branch b1) {$\bullet$}; 

      \draw[->] (rsc branch b2.center) to (rsc node1);
      \draw[->] (rsc branch b1.center) to (rsc node2);
    \end{scope}

\end{tikzpicture}
\label{subfig:ttcm_rsc}}

\subfloat[][NB-FEC Decoder]{
\begin{tikzpicture}

\begin{scope}

\coordinate (channel);

    \node[right=20pt of channel, draw, fill=Set3-8-7] (demapper) {$\matinv{M}$};
    \draw[->] (channel) -- (demapper) node[pos=0.4,label={[xshift=0em,yshift=-.5em]$\vect{y}$}] (y) {$/^1$};

    \node[above right= 28pt and 18pt of demapper, draw, fill=Set3-8-6] (decoder1) {BCJR$_1$};
    \node[right=12pt of decoder1,draw,circle,inner sep=0,fill=white] (add1) {$+$};
    \draw[->] (decoder1) -- (add1);

    \draw[->] (demapper) -| node[pos=0.3] {$/^8$} ($(decoder1.west)+(-4pt,0)$)  ($(decoder1.west)+(-4pt,0)$) -- (decoder1.west);

    \coordinate[below=5pt of decoder1.south] (extr1);
    \fill (extr1) circle[radius=1pt,anchor=mid];
    \draw[->] (extr1) -| node[near end,xshift=2pt] {$-$} (add1) node[pos=.3] {$/^2$}; 
    \draw[->] (extr1) -- (decoder1);

    \node[below right= 32pt and 18pt of demapper, draw, fill=Set3-8-6] (decoder2) {BCJR$_2$};
    \node[right=12pt of decoder2,draw,circle,inner sep=0,fill=white] (add2) {$+$};
    \draw[->] (decoder2) -- (add2) node[midway,fill,circle,inner sep=1pt] (out) {};

    \draw[->] (demapper) -| ($(decoder2.west)+(-4pt,0)$)  ($(decoder2.west)+(-4pt,0)$) -- (decoder2.west);

    \coordinate[above=5pt of decoder2.north] (extr2);
    \fill (extr2) circle[radius=1pt,anchor=mid];
    \draw[->] (extr2) -| node[near end,xshift=2pt] {$-$} (add2) node[pos=.3] {$/^2$}; 
    \draw[->] (extr2) -- (decoder2);

    \draw[-] (add1) -| ($(add1)+(8pt,-12pt)$) to[out=270,in=90] node[inner sep=5pt,pos=0.2,draw,fill=Set3-12-12] {$\Pi_\mathrm{s}$} (extr2) node[pos=0.2,right] {$L_\mathrm{e}$};
    \draw[-] (add2) -| ($(add2)+(8pt, 12pt)$) to[out=90,in=270] node[inner sep=5pt,pos=0.2,draw,fill=Set3-12-12] (deint) {$\Pi_\mathrm{s}^{\scriptscriptstyle-\!1}$}   node[pos=0.9,left] {~$L_\mathrm{a}$} (extr1);

\coordinate  (decoder2 input) at ($(decoder2.west)+(-4pt,0)$);

	\node[inner sep=5pt,draw,fill=Set3-12-12]  at (decoder2 input|-deint) {$\Pi_\mathrm{s}$};

    \node[ below right=12pt and 48pt of out,draw,align=left, fill=white] (sink) {Hard\\Decision};
    \draw[->] (out) |- (sink) node[inner sep=5pt,pos=0.65,draw,fill=Set3-12-12] {$\Pi_\mathrm{s}^{\scriptscriptstyle-\!1}$} node[pos=.9,label={[xshift=0,yshift=-.5em]$\vect{\lambda}$}] {$/^2$};
\draw[->] (sink.east) -- +(24pt,0) node [pos=.6,label={[yshift=-.5em]above:$\vect{\hat{c}}$}] (c) {$/^2$};


 \begin{scope}[on background layer]  
\node[draw,fit=(demapper)(decoder2)(decoder1)(sink),inner sep=4pt,fill=Set3-8-5] (box) {};
\end{scope}


    \end{scope}
\end{tikzpicture}
\label{subfig:ttcm_dec}}

  \caption[TTCM Encoder and decoder]{%
The TTCM encoder \subref{subfig:ttcm_enc}, and decoder \subref{subfig:ttcm_dec}.
The encoder consists of two $R=2/3$ RSC encoders \subref{subfig:ttcm_rsc}.
The second one works on the interleaved ($\Pi_s$) bits and its output is immediately deinterleaved ($\matinv{\Pi_s}$).
The encoder outputs are then punctured and mapped to 8PSK symbols for transmission.
The decoder implements a symbol-wise soft demapper and then the odd and even symbols are split and sent into two BCJR decoders that pass only soft information on data bits to each other.}
  \label{fig:ttcm_enc}
\end{figure}

The \ac{ttcm} scheme we consider in this paper was introduced by Robertson and Worz in \cite{Robertson1998} and is shown in Fig.~\ref{fig:ttcm_enc}.
In this scheme, two $R=2/3$ \ac{rsc} encoders,  with 8 states, encode the same data bits.
The encoder structure is shown in Fig.~\ref{fig:ttcm_enc}~\subref{subfig:ttcm_rsc}, where $Z$ are delay elements and the additions are modulo-2.
The symbol-wise encoders work on pairs of data bits to create a 3-bit symbols containing 2 data bits and one parity bit.
One of the encoders ($\mathrm{RSC}_1$) works directly on the 2-bit symbols, while the second encoder ($\mathrm{RSC}_2$) works on symbol-wise interleaved ($\Pi_\mathrm{s}$) 2-bit symbols (see Fig.~\ref{fig:ttcm_enc}).
The output of the second encoder is then symbol-wise de-interleaved ($\matinv{\Pi_\mathrm{s}}$) to realign the parity bit from this encoder with the original data bit pairs.
The encoded 3-bit symbols are then punctured, such that output symbols consist of the odd symbols from the first encoder and the even symbols from second encoder. The 3-bit symbols are then mapped to \ac{8psk} symbols using a natural binary mapping.
The symbol-wise interleaver is random and has the constraints that it maps odd to odd and is ``s-random'' to ensure that the corresponding trellis diagram has no parallel transitions \cite{Dolinar1995WeightPermutations}.

At the receiver, shown in Fig.~\ref{fig:ttcm_enc}~\subref{subfig:ttcm_dec}, the received symbols $\vect{y}$ are converted into 8 \acp{ll} by the symbol-wise soft demapper $\matinv{M}$. Because the odd symbols are produced by $\mathrm{RSC}_1$ and the even symbols by $\mathrm{RSC}_2$, at the receiver we then separate the odd and even symbols to send these to separate decoders. The two decoders are based on the \ac{bcjr} \cite{Bahl1974OptimalCorresp.} algorithm and work independently of each other by interchanging only soft information on the data bits.
The first decoder ($\mathrm{BCJR}_1$) works on the \acp{ll} from the odd symbols, where the \acp{ll} from even symbols are substituted by $0$. 
The first decoder also uses \emph{a priori} information on the data bits provided by the second decoder ($\mathrm{BCJR}_2$).
The \emph{a priori} information ($L_a$) is subtracted from the output of the first decoder to obtain the extrinsic information ($L_e$) which is then passed to the second decoder.
The second decoder works on the \acp{ll} from even symbols, substituting the odd symbol \acp{ll} by $0$.
At the first iteration---and following \cite{Robertson1998}---the \emph{a priori} information at the first decoder is given by a special metric. This metric is calculated by taking sum of the \acp{ll} of the symbols whose data bits are identical. This metric is only calculated at the positions of the even symbols; at the odd positions, zeros are used. The two decoders are then run sequentially for 10 iterations passing extrinsic information at each iteration. We chose 10 iterations because that number resulted in a decoder performance within $0.1$~dB of the best achievable performance (obtained with 100 iterations). 


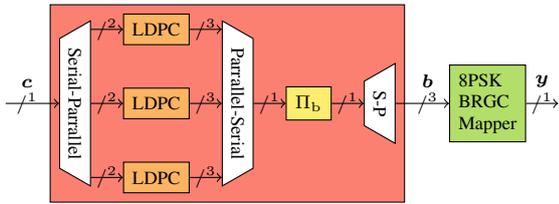
\begin{figure}[tbp]
  \centering
  \begin{tikzpicture}[every node/.style={node distance=12pt,font=\scriptsize}]


\coordinate (origin);

\node[right=20pt of origin,draw,trapezium, shape border rotate=90, fill=white] (sp1) {\rotatebox{-90}{Serial-Parrallel}};
\draw[->] (origin) -- (sp1) node[pos=0.4,label={[xshift=0,yshift=-.5em]above:$\vect{c}$}] (data bits) {$/^1$};


\node[right=of sp1, draw, fill=Set3-8-6] (enc mid) {LDPC};
\draw[->] (sp1) -- (enc mid) node[midway] {$/^2$};

\node[above=16pt of enc mid,draw, fill=Set3-8-6] (enc top) {LDPC};
\draw[->] (enc top -| sp1.east) -- (enc top) node[midway] {$/^2$};

\node[below=16pt of enc mid,draw, fill=Set3-8-6] (enc bot) {LDPC};
\draw[->] (enc bot -| sp1.east) -- (enc bot) node[midway] {$/^2$};


\node[right=of enc mid, draw, trapezium, shape border rotate=270, fill=white] (ps) {\rotatebox{-90}{Parrallel-Serial}};

\draw[->] (enc mid) -- (ps) node[midway] {$/^3$};
\draw[->] (enc top) -- (enc top -| ps.west) node[midway] {$/^3$};
\draw[->] (enc bot) -- (enc bot -| ps.west) node[midway] {$/^3$};


\node[right=of ps,draw, fill=Set3-12-12] (interleaver) {$\Pi_\mathrm{b}$};
\draw[->] (ps) -- (interleaver) node[midway] {$/^1$};


\node[right=of interleaver, draw, trapezium, shape border rotate=90, fill=white] (sp2) {\rotatebox{-90}{S-P}};
\draw[->] (interleaver) -- (sp2) node[midway] {$/^1$};

\node[right=20pt of sp2, draw,align=left, fill=Set3-8-7] (mapper) {8PSK \\ BRGC \\ Mapper};

\draw[->] (sp2) -- (mapper) node[pos=.6, label={[yshift=-.5em]above:$\vect{b}$}] {$/^3$};

\coordinate[right=of mapper] (symbols);

\draw[->] (mapper) -- (symbols) node[midway, label={[xshift=0,yshift=-.5em]above:$\vect{y}$}] {$/^1$};

\begin{scope}[on background layer]

\node[draw,fit=(sp1)(enc top)(enc bot)(sp2), fill=Set3-8-4] {};

\end{scope}

  \end{tikzpicture}
  \caption{LDPC encoder implementation. The incoming data bits are sequentially deserialised into 3 separate 2 bit wide streams and fed into independent identical rate 2/3 LDPC encoders. The 3 bit wide outputs are then serialized and then bit wise interleaved before mapping 8PSK symbols using a BRGC.}
  \label{fig:ldpc_enc}
\end{figure}

\subsection{\ac{bicm} \ac{ldpc}}
The \ac{ldpc} coding scheme we consider in this paper is based on the \ac{ldpc} from the DVB-S2 standard~\cite{Standard2013Modulation}.
The employed encoder structure is depicted in Fig.~\ref{fig:ldpc_enc}.
The data bits are deserialised into three streams each two bits wide and sent to three identical rate $R=2/3$ \ac{ldpc} encoders each producing three output bits.
All the encoded bits are re-serialised and a bit-wise interleaver was then used to interleave the codewords of all three encoders.
The interleaved bits were then mapped to \ac{8psk} symbols using the \ac{brgc}.

The LDPC receiver is essentially the reverse of the encoder shown in Fig.~\ref{fig:ldpc_enc}. Similarly to the \ac{ttcm} receiver, the symbols are first soft demapped into 8 \acp{ll} corresponding to the \ac{8psk} symbols and then, unlike in the TTCM receiver, the symbol-wise \acp{ll} are converted into 3 \emph{bit-wise} \acp{llr}. The \acp{llr} are then de-interleaved and split into the three different \ac{ldpc} codewords. After 50 iterations of the \ac{ldpc} decoder, the performance of the system was assessed. The 50 iterations were also chosen such that the performance was within $0.1$~dB of the asymptotic performance.

\subsection{Codeword length}

Throughout this paper we use $N_\mathrm{s}$ to denote the number of symbols in the transmitted codeword, i.e., $\underline{\vect{x}}=[\vect{x}^{(1)},\vect{x}^{(2)},\ldots,\vect{x}^{(N_\mathrm{s})}]$.\footnote{Throughout this paper we use the boldface letters (e.g., $\vect{x}$) to denote real (column) vectors and underlined boldface letters (e.g., $\underline{\vect{x}}$) to denote sequences of vectors.}
In this section we study the impact of the codeword length $N_\mathrm{s}$ on the performance of both \ac{cm} schemes. This was done to ensure that chosen values of $N_\mathrm{s}$ did not have a significant impact on the obtained results.
We will consider $N_\mathrm{s}=64800$ and $N_\mathrm{s}=21600$.

For the \ac{ttcm} scheme of Fig.~\ref{fig:ttcm_enc} that operates on symbols, the codeword length is the length of the interleaver.
For the \ac{bicm} \ac{ldpc} scheme of Fig.~\ref{fig:ldpc_enc}, the individual encoders produce bit sequences of length $64800$, which after serialisation and interleaving gives $N_\mathrm{s}=64800$. To generate $N_s=21600$ symbols, only one of the LDPC encoders was used and the interleaver was omitted.

\begin{figure}[tpb]
\centering
\begin{tikzpicture}
\begin{semilogyaxis}[
	xlabel={SNR [dB]},
    ylabel={$\log_{10}\left( \mathrm{BER} \right)$},
    xmin=5.5,
    xmax=7.5,
    ymode=log,
    log base 10 number format code/.code={\pgfmathprintnumber[fixed]{#1}},
	grid=major,
    legend style={
    	at={(0.5,1.03)},
        anchor=south,
    },
    legend cell align=left,
    legend columns=2,
	cycle multi list={%
		Set1-4\nextlist
    	[2 of]mark list
	},
]

\addplot +[Paired-12-1] table [x=snr, y=ber, col sep=comma] {data/ttcm_2p16_th.csv};
\addlegendentry{TTCM $N_\mathrm{s}=64800$ Symbols};
\addplot +[Paired-12-5,mark options={fill=white}] table [x=snr, y=ber, col sep=comma] {data/ldpc_64800_th.csv};
\addlegendentry{LDPC $N_\mathrm{s}=64800$ Symbols};
\addplot +[Paired-12-2] table [x=snr, y=ber, col sep=comma] {data/ttcm_21600_th.csv};
\addlegendentry{TTCM $N_\mathrm{s}=21600$ Symbols}
\addplot +[Paired-12-6] table [x=snr, y=ber, col sep=comma] {data/ldpc_21600_th.csv};
\addlegendentry{LDPC $N_\mathrm{s}=21600$ Symbols};

\end{semilogyaxis}
\end{tikzpicture}
\caption{\ac{ber} performance of \ac{ttcm} and \ac{bicm} \ac{ldpc} with different codeword lengths.}
\label{fig:length_comp}
\end{figure}
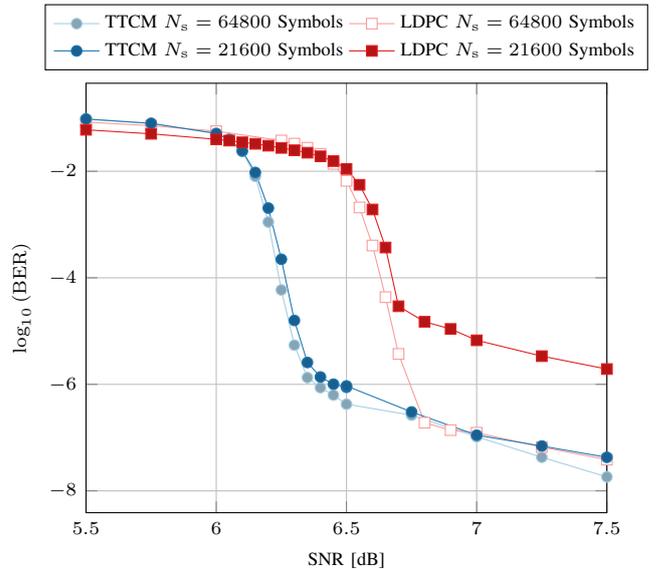

Fig.~\ref{fig:length_comp} shows the impact of reducing the codeword length from $N_\mathrm{s}=64800$ to $N_\mathrm{s}=21600$ symbols on the post-\ac{fec} \ac{ber} performance.
These results were obtained using an \ac{awgn} channel and show that the impact of codeword length $N_\mathrm{s}$ to both schemes is minimal in the convergence region. However, a longer codeword length reduces the error floor for the \ac{ldpc} scheme. 
{Since a threefold increase in codeword length only delivered minor improvements, which indicates this is the in the convergence region, increasing the codeword length even further will only deliver diminishing returns. Furthermore, using $N_\mathrm{s}>64800$ makes the post-\ac{fec} \ac{ber} below the hard decision \ac{fec} threshold of $5\cdot10^{-5}$ for a $1\%$ overhead \ac{fec} \cite{Sugihara2013AGbps}.}
Therefore, from now on, we use a codeword length of $N_\mathrm{s}=64800$ symbols. When $N_\mathrm{s}=64800$ symbols, the results in Fig.~\ref{fig:length_comp} also show how \ac{ttcm} outperforms \ac{bicm} \ac{ldpc} by about $0.5$~dB.

\section{Performance measures}
\label{sec:method}

\newcommand{\condsumn}{\frac1{|\mathcal{N}_i|}\sum_{n\in\mathcal{N}_i}}

At the receiver side of the bit-wise receiver in Fig.~\ref{fig:intro}~\subref{subfig:intro_bicm}, the decoder works on soft information available on the encoded bits.
In such a system, the most popular performance metric is the pre-\ac{fec} \ac{ber}, computed after hard-decision demapping, or equivalently after hard-decisions on the \acp{llr} $\vect{l}=[l_1,\ldots,l_m]^\mathrm{T}$.
{This metric might be an accurate predictor of the performance of coded modulation for small constellation sizes and \acp{snr}, however, its use has no theoretical foundation. Furthermore, this metric is in general a poor predictor of performance of coded modulation, as shown in \cite{Alvarado2015ReplacingSystems,Schmalen2016PredictingExperiments,2016arXiv160600755S}.}
Furthermore, when considering the symbol-wise decoder in Fig.~\ref{fig:intro}~\subref{subfig:intro_ttcm}, the encoded bits are completely absent at the receiver, and hence, pre-\ac{fec}-\ac{ber} cannot be used either \cite{Schmalen2016PredictingExperiments,2016arXiv160600755S}.
In this work, we use an information-theoretical approach and consider \acp{air}. In particular---following \cite{Alvarado2015ReplacingSystems} and  \cite{Schmalen2016PredictingExperiments,2016arXiv160600755S}---we consider \ac{mi} and \ac{gmi} to assess and compare the performance of the two systems under consideration. Furthermore, we will also consider the post-\ac{fec} \ac{mi} as a way to estimate the ultimate performance of the system considering the BICM LDPC and TTCM decoders.

The channel is modelled as a multi-dimensional correlated real memoryless channel $\vect{Y} =\vect{X}+\vect{Z}$ with transmitted symbols $\vect{x}=[x_1,x_2,\ldots,x_{N_\mathrm{D}}]^\mathrm{T}\in \mathcal{X}\subset \mathbb{R}^{N_\mathrm{D}}$, additive noise $\vect{z}=[z_1,z_2,\ldots,z_{N_\mathrm{D}}]^\mathrm{T} \in \mathbb{R}^{N_\mathrm{D}}$ and received symbols $\vect{y}=[y_1,y_2,\ldots,y_{N_\mathrm{D}}]^\mathrm{T} \in \mathbb{R}^{N_\mathrm{D}}$.\footnote{In this paper, a real-valued random vector is denote by $\vect{X}$ and its outcome by $\vect{x}$. Sets are denoted by calligraphic letters $\mathcal{X}$ and $[\cdot]^\mathrm{T}$ denotes transpose. The norm of a vector $\vect{x}$ is denoted by $\| \vect{x} \|$. {$\mathbf{\Sigma}$ is a matrix.}}
Here, $\mathcal{X}=\{\vect{x}_1,\vect{x}_2,\ldots,\vect{x}_M\}$ is the set of constellation points, where $|\mathcal{X}|=M=2^m$ and $N_\mathrm{D}$ is the number of dimensions.
The channel transition probability is given by%
{\begin{multline}
 f_{\vect{\vect{Y}|\vect{X}}}(\vect{y}|\vect{x}) = \frac{1}{\sqrt{(2\pi)^{N_\mathrm{D}} \det(\mathbf{\Sigma})}}   
 \\ \exp \left( -\frac12 (\vect{y}-\vect{x})^\mathrm{T} \mathbf{\Sigma}^{-1} (\vect{y}-\vect{x}) \right)\label{eq:2d_g}
\end{multline}}
where {$\mathbf{\Sigma}$} is the covariance matrix.

In this paper, we consider the model in \eqref{eq:2d_g}, because it has been previously shown to accurately model the noise from the optical transmission \cite{Eriksson2016ImpactExperiments}. Furthermore, this model allows us to better describe phase noise acquired during transmission due to the Kerr effect.
In all the performance assessments presented in this paper, we will sweep the \ac{snr} which is defined as $\mathrm{SNR}=  \mathbb{E}\left[ \lVert\vect{X}\rVert^2 \right] / \mathbb{E}\left[ \lVert\vect{Z}\rVert^2 \right]$, where $\mathbb{E}[\cdot]$ denotes expectation. 

As mentioned before, in this work we will use three information-theoretic performance measures: \ac{mi}, \ac{gmi}, and post--\ac{fec} \ac{mi}. \ac{mi} is an \ac{air} for \ac{cm} based on \ac{nbfec} (e.g., for \ac{ttcm} and also for \ac{bicm} with iterative demapping and decoding) and \ac{gmi} is an \ac{air} for \ac{cm} based on binary \ac{fec} and bit-wise decoding (e.g., for the \ac{ldpc} scheme we consider in this paper). The post-\ac{fec} \ac{mi} is an \ac{air} for an outer code used after the CM decoder.
{Please note that the \acp{air} used in this paper are a lower bound on the AIRs of the true channel due to the mismatch between the chosen channel law and the true channel law \cite{Colavolpe2011FasterInterference,Secondini2013AchievableMaps,Ganti2000MismatchedLimit,Arnold2006SimulationMemory}.}
In the following sections, we derive a closed form expression to approximate the \ac{mi} and \ac{gmi} using channel observations and show an expression to calculate the post-\ac{fec} \ac{mi} using bit-wise \acsp{llr}.

\subsection{Mutual information}

The mutual information is defined as \cite{Shannon1948ACommunications}%
\begin{equation}
  I(\vect{X};\vect{Y})=\mathbb{E}\left[\log_2 \frac{f_{\vect{Y}|\vect{X}}(\vect{Y}|\vect{X})}{f_{\vect{Y}}(\vect{Y})} \right]\label{eq:mi_bound}
\end{equation}%
where $f_{\vect{Y}|\vect{X}}(\vect{y}|\vect{x})$ is the channel transition probability. In this paper we consider the correlated \ac{awgn} \ac{pdf} given by (\ref{eq:2d_g}) {and we will use a ready-to-use closed-form approximation for the \ac{mi} of this channel (shown below)}.
For a sequence of transmitted symbols $\vect{x}^{(n)}$ and received symbols $\vect{y}^{(n)}$ with $n=1,2,\ldots,N_s$, the \ac{mi} for the channel in \eqref{eq:mi_bound} can be approximated as
{\begin{multline}
	I(\vect{X};\vect{Y}) \approx\, m -\frac{1}{M} \sum^M_{i=1}\frac1{|\mathcal{N}_i|}  \\
  \sum_{n\in\mathcal{N}_i} \log_2  \sum_{j=1}^{M} \exp \left(-\frac12 \vect{d}_{ij}^\mathrm{T} \matinv{\mathbf{\Sigma}} \vect{d}_{ij} - \vect{d}_{ij}^\mathrm{T} \matinv{\mathbf{\Sigma}} \vect{z}^{(n)} \right) \label{eq:mi}
\end{multline}}%
where 
\begin{equation}\label{Ni.set}
\mathcal{N}_i=\{n=1,2,\ldots,N_\mathrm{s}:\vect{x}^{(n)}=\vect{x}_i \}
\end{equation}
is the set of all timeslots where the $i$th constellation point was sent, $\vect{z}^{(n)}=\vect{y}^{(n)}-\vect{x}^{(n)}$, $\vect{d}_{ij}=\vect{x}_i-\vect{x}_j$, and {$\mathbf{\Sigma}$} is the covariance matrix.
{The derivation for this expression can be found in the Appendix.}


\subsection{Generalized mutual information}

 The \ac{gmi} \cite[eq.~(10)]{Martinez2009Bit-InterleavedPerspective} is defined as the sum of the mutual information between the encoded bits ($B_k$) and the received symbols ($\vect{Y}$),%
\begin{align}
\mathrm{GMI} &= \sum_{k=1}^m I(B_k;\vect{Y}) \label{eq:gmi_general}\\
&= \sum_{k=1}^m \mathbb{E}\left[\log_2 \frac{f_{\vect{Y}|B_k}(\vect{Y}|B_k)}{f_{\vect{Y}}(\vect{Y})} \right].\label{eq:gmi_analytical}
\end{align}
The following {expression} gives a closed-form approximation for the \ac{gmi} of a correlated \ac{awgn} channel.
For a sequence of transmitted symbols $\vect{x}^{(n)}$ and received symbols $\vect{y}^{(n)}$ with $n=1,2,\ldots,N_s$, the \ac{gmi} for the channel in \eqref{eq:gmi_analytical} can be approximated as %
{\begin{multline}
 \mathrm{GMI} \approx \,  m - \frac{1}{M}\sum^m_{k=1} \sum_{l\in\lbrace0,1\rbrace} \sum_{i\in\mathcal{I}_{l,k}}\frac1{|\mathcal{N}_i|}  \\
  \sum_{n\in\mathcal{N}_i} \log_2 \frac{\sum^M_{p=1} \exp \left(- \frac12 \vect{d}_{ip}^\mathrm{T} \matinv{\mathbf{\Sigma}} \vect{d}_{ip} - \vect{d}_{ip}^\mathrm{T} \matinv{\mathbf{\Sigma}} \vect{z}^{(n)} \right)}{\sum_{j\in\mathcal{I}_{l,k}} \exp \left(- \frac12 \vect{d}_{ij}^\mathrm{T} \matinv{\mathbf{\Sigma}} \vect{d}_{ij} - \vect{d}_{ij}^\mathrm{T} \matinv{\mathbf{\Sigma}} \vect{z}^{(n)} \right)}
\label{eq:gmi}
\end{multline}}
where 
\begin{equation}\label{Nlk.set}
\mathcal{I}_{l,k}=\{i=1,2,\ldots,M:\Phi(\vect{b})=\vect{x}_i,b_k=l\}
\end{equation}
is the set of indices of constellation points where the $k$th encoded bit in $\vect{b}=[b_1,b_2,\ldots,b_m]^\mathrm{T}$ has the value $l$ and $|\mathcal{I}_{l,k}|=M/2$.
{The derivation of \eqref{eq:gmi} can be found in the Appendix.}

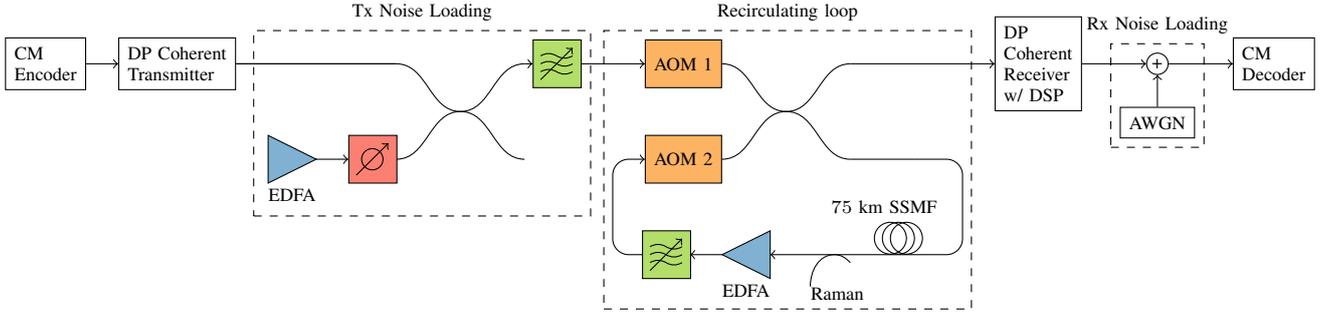
\begin{figure*}[tbp]
\centering
\begin{tikzpicture}[
block/.style={draw},
node distance= 12pt,
every node/.style={
font=\scriptsize,
}
]

\node[block, align=left] (source) {DP Coherent \\ Transmitter};

\node[block, left= of source, align=left] (encoder) {CM \\ Encoder};
\draw[->] (encoder) -- (source);



\path coordinate[right=of source] (helper);

\coordinate[right=60pt of source] (helper);
\draw[-] (source) -- (helper) pic (coupler) {Coupler};

\draw[-] (source) -- (coupler-in1);

\path coordinate[left=0 of coupler-in2] (helper);
\draw[-] (coupler-in2) -- (helper) pic[pic shift={(-18pt,0)}, pic fill=Set3-8-4] (att) {Attenuator};

\path coordinate[left=of att-in] (helper);
\draw[<-] (att-in) -- (helper) pic[pic shift={(-18pt,0)},pic fill=Set3-8-5] (edfa) {EDFA};

\coordinate[right=3pt of coupler-out1] (helper);
\draw[->] (coupler-out1) -- (helper) pic[pic fill=Set3-8-7] (filter) {Filter};

 \node[draw, dashed, fit=(att-box)(filter-box)(edfa-label)] (noiseloadbox) {};
\node[above=0 of noiseloadbox.north] {Tx Noise Loading};

\coordinate[right= of filter-out] (helper);
\draw[-] (filter-out) -- (helper) pic (loop) {Loop};

\node[right=of loop-out, draw, align=left] (receiver) {DP\\Coherent\\Receiver\\w/ DSP};

\draw[->] (loop-out) --(receiver);

\coordinate[right= of receiver] (rx helper);
\node[draw,circle, right= of rx helper, inner sep = 1pt] (rx noiseadd) {+};
\node[block,below= of rx noiseadd, fill=white] (rx noise) {AWGN};
\draw[->] (receiver) -- (rx noiseadd);
\draw[->] (rx noise) -- (rx noiseadd);

\node[draw, dashed,fit =(rx noiseadd)(rx noise)] (rx noisebox) {};
\node[above=0 of rx noisebox.north] {Rx Noise Loading};

\coordinate[right= of rx noiseadd] (dec helper);
\node[block, right= of dec helper, align=left] (decoder) {CM \\ Decoder};
\draw[->] (rx noiseadd) -- (decoder);

\end{tikzpicture}
\caption{The setup as used for the experimental results. First data bits are encoded into 8PSK symbols that are modulated at $28$~Gbaud onto $1550$~nm. Optionally, extra ASE is added to the signal for transmitter noise loading. A $1000$~km transmission is emulated using a recirculating loop. The signal is then received and is processed off-line. Optionally extra noise is loaded at the receiver before the codeword is decoded.}
\label{fig:setup}
\end{figure*}

Expressions similar to \eqref{eq:mi} and \eqref{eq:gmi} have been used in previous works.
For example, approximations for \ac{mi} and \ac{gmi} for a \ac{cs} \ac{awgn} multi-dimensional channel based on Gauss--Hermite quadrature were derived in \cite[Sec.~III]{Alvarado2011HighCapacity} (see also \cite[Sec.~4.5]{Szczecinski2015Bit-InterleavedDesign}).
Very recently, a generic Monte Carlo approximation for the multi-dimensional \ac{mi} was presented in \cite[eq.~(4)]{Eriksson2016MultidimensionalCommunications}.
To the best of our knowledge, however, {expressions \eqref{eq:mi} and \eqref{eq:gmi}} are the first to present closed-form approximations for the \ac{mi} and \ac{gmi} of constellations for a multi-dimensional correlated \ac{awgn} channel {and evaluate their use with experimentally obtained results}.

\subsection{Post-\ac{fec} mutual information}

\ac{cm} is typically designed to be combined with a low rate outer code to get the \ac{ber} down to the desired level (usually $10^{-15}$). 
In this section, we discuss two \acp{air} for this outer code, one for \ac{hd} codes, and one for \ac{sd} codes.
The relevance of these metrics is that when compared to the \ac{mi} and \ac{gmi}, they allow us to visualize the suboptimality of particular \ac{cm} implementations.

Both \ac{bicm} \ac{ldpc} and \ac{ttcm} decoders produce soft information on the data bits.
We denote this soft information as $\underline{\vect{\lambda}}=[\vect{\lambda^{(1)}},\vect{\lambda^{(2)}},\ldots,\vect{\lambda^{(N_\mathrm{s})}}]$
, where
\begin{equation}
\lambda_q^{(n)}=\log \left( \frac{f_{C_q^{(n)}|\underline{\vect{Y}}}(0|\underline{\vect{y}})}{f_{C_q^{(n)}|\underline{\vect{Y}}}(1|\underline{\vect{y}})} \right), \quad q={1,2} \label{eq:postfecllr}
\end{equation}
{are the \ac{llr} of the $q$th data bit at the $n$th symbol given the sequence of received symbols $\underline{\vect{y}}$. Note that this expression depends on the sequence of received symbols. This is because the decoder can use the whole sequence of received symbols to determine the bit probability. The fact that the decoder uses all the received symbols, however, does not imply that the channel has memory.}

As shown in Fig.~\ref{fig:ttcm_enc}~\subref{subfig:ttcm_dec} (for \ac{ttcm}) this soft information can be converted into (hard) bits, which we denote by $\hat{c}_1^{(n)}$ and  $\hat{c}_2^{(n)}$.

When the outer code is \ac{sd}, the information-theoretical quantity we consider is the post-\ac{fec} \ac{mi}, which is defined as
\begin{equation}
\mathrm{I}_\mathrm{SD} = \sum_{q=1}^2 I(C_q;\Lambda_q) \label{eq:pfmi_sd_theoretical}
\end{equation}
where $C_q$ and $\Lambda_q$ are the random variable which describes the data bits and the \acp{llr} $\lambda_q^{(n)}$, respectively. 
The \ac{mi} in \eqref{eq:pfmi_sd_theoretical} can be approximated as
\begin{equation}
\mathrm{I}_\mathrm{SD} \approx \sum^2_{q=1}\left(1 - \frac{1}{N_\mathrm{s}} \sum_{n=1}^{N_\mathrm{s}} \log_2 \left[ 1+\exp{\left( (-1)^{c_q^{(n)}} \lambda_q^{(n)}  \right)} \right]\right) \label{eq:postsdmi}
\end{equation}%
where $\underline{\vect{c}}=[\vect{c}^{(1)},\vect{c}^{(2)},\ldots,\vect{c}^{(N_\mathrm{s})}]$ and $\vect{c}=[c_1,c_2]^\mathrm{T}$.
The approximation in \eqref{eq:postsdmi} is obtained by assuming that the \ac{pdf} of the \acp{llr} in \eqref{eq:postfecllr} satisfies the so-called consistency condition \cite[Def.~3.8]{Szczecinski2015Bit-InterleavedDesign}, \cite[eq.~(12)]{Brink2001ConvergenceCodes} and by using a Monte Carlo approximation of the one-dimensional integral.
Note that under certain assumptions, the expression in \eqref{eq:postsdmi} can also be used to approximate the \ac{gmi} in \eqref{eq:gmi_general}.
This can be done by using \acp{llr} on encoded bits $\underline{\vect{b}}$ instead of data bits $\underline{\vect{c}}$, as done in \cite[eq.~(2)]{Fehenberger2016ImprovedExperiments} and \cite[eq.~(30)]{Alvarado2015ReplacingSystems}.

When the outer code is \ac{hd}, we consider the \ac{mi} between the information bits $C_1$ and  $C_2$ and their respective \ac{hd} estimates after decoding, i.e., 
\begin{align}
\mathrm{I}_{\mathrm{HD}} &= \sum_{q=1}^2I(C_q,\hat{C_q}) \\
&=\sum_{q=1}^2(1-H_q(\mathrm{BER}_q)) \label{eq:posthdmi}
\end{align}
where $\mathrm{BER}_q$ is the \ac{ber} at the $q$th decoder output and 
\begin{align}
H_b(p)&=-p\log(p)-(1-p)\log(1-p)
\end{align}
is the binary entropy function. Because of the data processing inequality, $\mathrm{I}_{\mathrm{SD}}\geq \mathrm{I}_{\mathrm{HD}}$.

\section{Experimental setup}
\label{sec:setup}

The experimental transmission setup is shown Fig.~\ref{fig:setup}. An {\acl{ecl}} at $1550$~nm is modulated by a {\acl{mzm}} driven by {an arbitrary waveform generator} at $28$~Gbaud for both in-phase and quadrature. 
Polarisation multiplexing is emulated by splitting the signal into two identical single polarisation signals, delaying one of the two signals and then recombining using a {\acl{pbs}}.
Transmitter-based noise loading was used to vary the \ac{snr}, by adding additional \ac{ase} from an \ac{edfa}. The signal was then transmitted using a recirculating loop. The recirculating loop consisted of a 75~km \ac{ssmf} span with both \ac{edfa} and Raman amplification and was used to emulate transmission over a total distance of $1000$~km. The launch power per span was set to 0~dBm to ensure linear propagation. A bandpass filter was used to remove the out-of-band noise and two \acp{aom} were used control the loading to the signal into the loop.

The signal is then received by a \ac{dp} coherent receiver. Standard off-line \ac{dsp} \cite{vanUden2014MIMOSystems.} was used to equalize the signals and recover the noisy \ac{8psk} symbols. The recovered constellations are then passed to the \ac{cm} decoder.

The transmitted sequences were generated by encoding identical pseudo-random bit sequences with either the \ac{ttcm} or the \ac{ldpc} encoder from Sec.~\ref{sec:cm}. Codewords consisting of $N_\mathrm{s}=64800$ \ac{8psk} symbols were transmitted and at the receiver, a single trace contained $7$ codewords in each polarization, yielding $2.7\times10^{6}$ encoded bits or $1.8\times10^{6}$ data bits after decoding.

To further investigate the decoding performance, receiver-based noise loading was also employed. This was implemented by obtaining an experimental trace without transmitter noise loading after transmission over 1000~km and then \ac{awgn} was added digitally to the recovered constellation before decoding. The effectiveness of noise loading at the receiver will be explained in the next section.

\begin{figure}[ptb!]
\centering
\begin{tikzpicture}
\begin{semilogyaxis}[
	width=\columnwidth,height=.75\columnwidth,
	xlabel={SNR [dB]},
    ylabel={$\log_{10}\left( \mathrm{BER} \right)$},
    xmin=5.5,
    xmax=7.5,
    ymode=log,
    ymin=.5E-9,
    ymax=.5,
    log base 10 number format code/.code={\pgfmathprintnumber[fixed]{#1}},
	grid=major,
    legend style={
    	at={(0.5,1.03)},
        anchor=south,
    },
    legend cell align=left,
    legend columns=2,
	cycle multi list={%
		Set1-4\nextlist
    	[2 of]mark list
	},
]
\addlegendimage{empty legend}
\addlegendentry{NB-FEC}
\addlegendimage{empty legend}
\addlegendentry{Binary-FEC}

\addplot +[Paired-12-1,mark=no] table [x=snr, y=ber, col sep=comma] {data/ttcm_2p16_th.csv};
\addlegendentry{TTCM AWGN};\label{pgf:ber_ttcm}
\addplot +[Paired-12-5,mark=no] table [x=snr, y=ber, col sep=comma] {data/ldpc_64800_th.csv};
\addlegendentry{LDPC AWGN};\label{pgf:ber_ldpc}



\addplot [Set1-4-2, only marks, mark=+] table [x=snrs, y=bers, col sep=comma] {data/ttcm_experimental_per_codeword.csv};
\addlegendentry{TTCM Tx Noise loaded};
\addplot [Set1-4-1, only marks, mark=+] table [x=snrs, y=bers, col sep=comma] {data/ldpc_experimental_per_codeword.csv};
\addlegendentry{LDPC Tx Noise loaded};

\addplot +[Paired-12-2,only marks,mark=square*] table [x expr=(\thisrow{snr_1}+\thisrow{snr_2})/2, y expr=(\thisrow{ber_1}+\thisrow{ber_2})/2, col sep=comma] {data/ttcm_art.csv};
\addlegendentry{TTCM Rx Noise loaded};
\addplot +[Paired-12-6,only marks,mark=square*, mark options={fill=white}] table [x expr=(\thisrow{snr_1}+\thisrow{snr_2})/2, y expr=(\thisrow{ber_1}+\thisrow{ber_2})/2, col sep=comma] {data/ldpc_art.csv};
\addlegendentry{LDPC Rx Noise loaded};

\addplot +[thick,Paired-12-3,mark=no] table [x=snr, y=cm, col sep=comma] {data/min_ber.csv};
\addlegendentry{Minimum BER NB-FEC};\label{pgf:ber_mi}
\addplot +[thick,Paired-12-7,mark=no] table [x=snr, y=bi, col sep=comma] {data/min_ber.csv};
\addlegendentry{Minimum BER Bin-FEC};\label{pgf:ber_gmi}

\draw[<->] (axis cs:5.75,0.5e-4) -- (axis cs:6.25,0.5e-4) node [midway,label=above:{0.5~dB}] {};
\draw[<->] (axis cs:5.83,0.2e-4) -- (axis cs:6.66,0.2e-4) node [pos=0.23,label=below:{0.8~dB}] {};

\draw[<->] (axis cs:5.758,1e-6) -- (axis cs:5.83,1e-6) node [midway,label=below:{0.1~dB}] {};

\node[Set1-4-2,draw,very thick,ellipse,minimum height=1cm,minimum width=.3cm,label=below:TTCM] at (axis cs:6.4,0.5e-6)  {};

\node[Set1-4-1,draw,very thick,ellipse,minimum height=.3cm,minimum width=1cm,label=right:{BICM LDPC}] at (axis cs:6.6,2e-3)  {};

\end{semilogyaxis}

\end{tikzpicture}
\caption{BER of the TTCM and BICM LDPC compared to the theoretical performance in an AWGN channel and the comparison of transmitter and receiver noise loading.}
\label{fig:ber_art}
\end{figure}
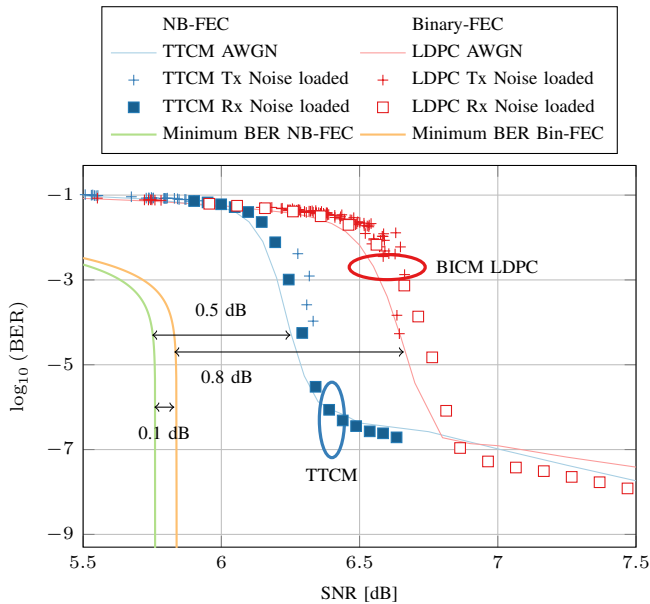

\begin{figure*}[tbp]
\centering
\begin{tikzpicture}[
	every tick label/.append style={font=\normalsize},
    every axis label/.append style={font=\normalsize}]
\begin{axis}
[	width=1.0\textwidth,
    height=1.2\columnwidth,
	xlabel={SNR [dB]},
    ylabel={AIR [bit/sym]},
    grid=major,
    xmin=4.2,
    xmax=7.2,
    ymin=1,
    ymax=2.4,
    ylabel style={yshift=-0.2cm},
    xlabel style={yshift=+0.1cm},
    legend style={font=\scriptsize,legend cell align=left,xshift=+.4cm,yshift=-.15cm},
    legend pos=south east,
    cycle multi list={%
		Set1-4\nextlist
    	[2 of]mark list
	},
]

\addplot [very thick, black,domain=-5:12,samples=200, restrict y to domain=0:3] {log2(1+10^(x/10))};
\addlegendentry{AWGN capacity};

\addplot [thick, Paired-12-3, mark=no] table [x=snr, y=mi, col sep=comma] {data/info8psk_brgc.csv};
\addlegendentry{MI AWGN};


\addplot [thin,  Paired-12-4, only marks, mark=triangle*, mark size=1.2pt] table [x=snr, y=condmi2d, col sep=comma] {data/calculation.csv};
\addlegendentry{MI Tx noise loaded};

\addplot [thin, Paired-12-7, mark=no] table [x=snr, y=gmi, col sep=comma] {data/info8psk_brgc.csv};
\addlegendentry{GMI AWGN};


\addplot [thick,  Paired-12-8, only marks, mark=triangle*, mark options={fill=white}, mark size=1.2pt, every mark/.append style={rotate=90}] table [x=snr, y=condgmi2d, col sep=comma] {data/calculation.csv};
\addlegendentry{GMI Tx noise loaded};


\addplot +[Paired-12-2, mark=no,mark size=1.25pt] table [x=snr, y=info, col sep=comma] {data/ttcm_2p16_th.csv};
\addlegendentry{TTCM AWGN SD};

\addplot +[Paired-12-1, only marks, mark=square*,mark size=1.25pt] table [x=snr, y=gmi, col sep=comma] {data/ttcm_info_full.csv};
\addlegendentry{TTCM Rx noise loaded SD};

\addplot +[Paired-12-10, mark=no,mark size=1.25pt] table [x=snr, y=hardinfo, col sep=comma] {data/ttcm_2p16_th.csv};
\addlegendentry{TTCM AWGN HD};
\addplot +[Paired-12-10, only marks, mark=*,mark size=1.25pt] table [x=snr, y=hb, col sep=comma] {data/postfec_hd_ttcm.csv};
\addlegendentry{TTCM Rx noise loaded HD};

\addplot +[Paired-12-6, mark=no,mark size=1.25pt] table [x=snr, y=info, col sep=comma] {data/ldpc_64800_th.csv};
\addlegendentry{LDPC AWGN SD};
\addplot +[Paired-12-5, only marks, mark=square*, mark options={fill=white},mark size=1.25pt] table [x=snr, y=gmi, col sep=comma] {data/ldpc_info_full.csv};
\addlegendentry{LDPC Rx noise loaded SD};

\addplot +[Paired-12-12, mark=no,mark size=1.25pt] table [x=snr, y=hardinfo, col sep=comma] {data/ldpc_64800_th.csv};
\addlegendentry{LDPC AWGN HD};
\addplot +[Paired-12-12, only marks, mark=*, mark options={fill=white},mark size=1.25pt] table [x=snr, y=hb, col sep=comma] {data/postfec_hd_ldpc.csv};
 \addlegendentry{LDPC Rx noise loaded HD};



\draw[<->] (axis cs:5.78,2.05) -- (axis cs:5.84,2.05) node [midway,above] {$0.06$~dB};
\draw[-] (axis cs:5.78,2) -- (axis cs:5.78,2.05) (axis cs:5.84,2.05) -- (axis cs:5.84,2);

\draw[<->] (axis cs:6.2,2.2) -- (axis cs:6.6,2.2)  node [midway,above] {$0.4$~dB};
\draw (axis cs:6.2,2) -- (axis cs:6.2,2.2) (axis cs:6.6,2.2) --(axis cs:6.6,2);

\node[draw,thick,ellipse,minimum height=2cm,minimum width=.75cm,label=above:SD] at (axis cs:5.4,1.35)  {};

\node[draw,thick,ellipse,minimum height=1.5cm,minimum width=.5cm,label=below:HD] at (axis cs:6,1.35)  {};

\draw[<-] (axis cs:5,1.82) -- + (.5cm,-.75cm) node[below] {GMI};

\draw[<-] (axis cs:5,1.87) -- + (.5cm,.75cm) node[above] {MI};

\node[draw,thick,ellipse,minimum height=.75cm,minimum width=1cm,label=left:TTCM] at (axis cs:6.15,1.9)  {};
\node[draw,thick,ellipse,minimum height=.75cm,minimum width=1cm,label=right:LDPC] at (axis cs:6.5,1.9)  {};

\end{axis}
\end{tikzpicture}
\caption{Performance of coding schemes in terms of \acp{air}. The \ac{awgn} capacity is shown as a reference.}
\label{fig:th_info}
\end{figure*}
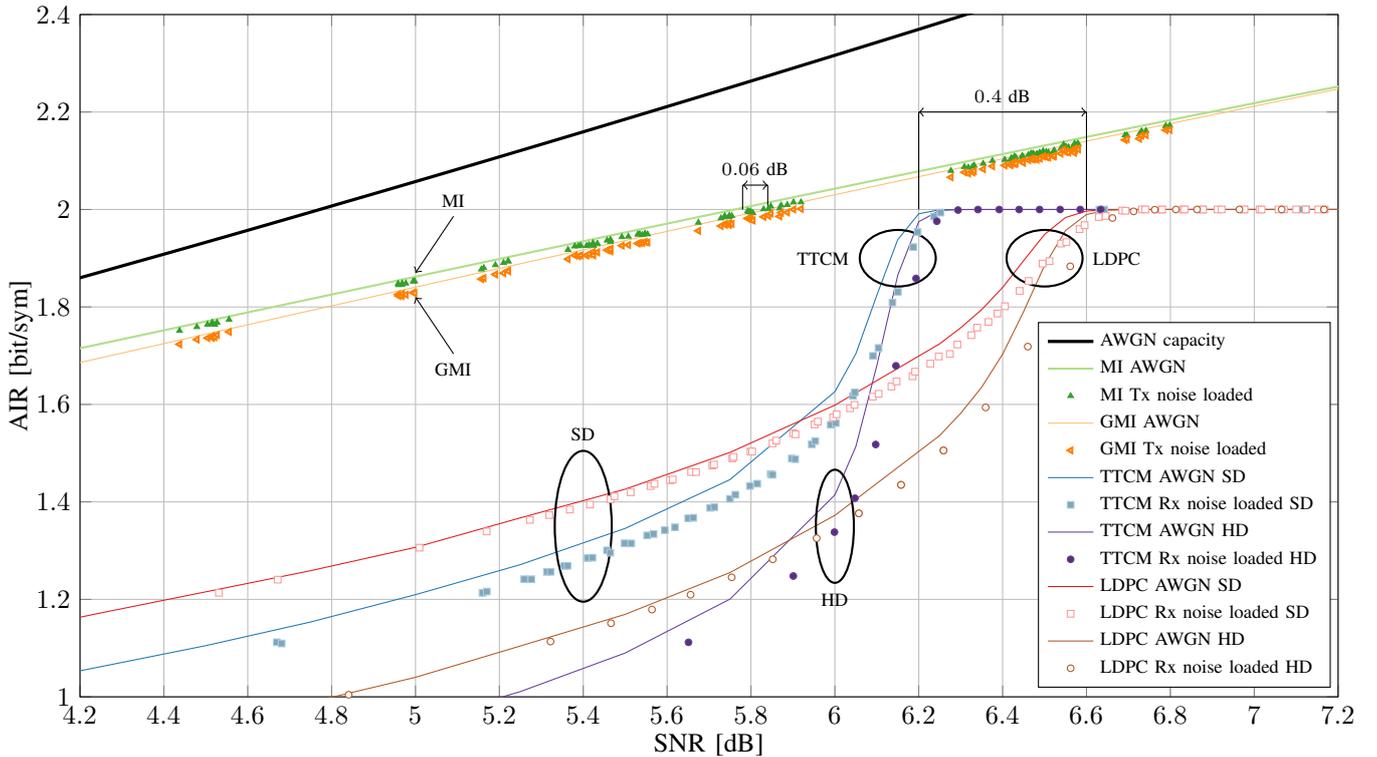

\section{Experimental results}
\label{sec:results}

Fig.~\ref{fig:ber_art} shows the post \ac{fec}-\ac{ber} performance of the two coding schemes as a function of \ac{snr}, where the \ac{snr} is measured from the recovered constellations at the receiver and the covariance matrix is estimated for each \ac{snr} value.  We find that the measured performance (markers) for both schemes matches well the calculated performance for the implemented schemes assuming an \ac{awgn} channel (thin lines). An implementation penalty of less than $0.1$~dB for both schemes is observed. We also see that both transmitter-based noise loading and receiver-based noise loading give very similar performance.

Fig.~\ref{fig:ber_art} also shows a theoretical lower bound on the \ac{ber} for \ac{8psk} and code rate $R=2/3$. This bound is also known as the rate distortion bound \cite{Leven2011EstimationExperiments}.
At this bound the binary entropy of \ac{ber} multiplied by the number of data bits matches the \ac{air} for the given \ac{snr}, i.e., $2(1-H_b(\mathrm{BER}))=\mathrm{AIR}$.
In the case of the \ac{ttcm}, the \ac{air} used in this equation is the \ac{mi} in \eqref{eq:mi_bound}. In the case of \ac{bicm} \ac{ldpc}, the \ac{air} used in the equation is the \ac{gmi} in \eqref{eq:gmi_analytical}.
The distance between the bound and the actual performance is the penalty incurred from design and implementation of the actual code.
We see that the implemented coding schemes are $0.5$ and $0.8$~dB away from the minimum theoretical lower bounds on \ac{ber} given by \ac{mi}(\ref{pgf:ber_mi}) and \ac{gmi}(\ref{pgf:ber_gmi}) respectively, for \ac{ttcm}(\ref{pgf:ber_ttcm}) and \ac{ldpc}(\ref{pgf:ber_ldpc}) respectively.
As we will see below, these different gaps to the theoretical bounds also appear when \acp{air} are considered.
{Note also that the losses of $0.5$ and $0.8$~dB are due to the particular choice of TTCM and LDPC codes we consider here. The gaps for better codes could be smaller, however, the $0.1$~dB gap given by the theoretical curves in Fig.~\ref{fig:ber_art} will always remain the same.}

For transmitter-based noise loading (crosses in Fig.~\ref{fig:ber_art}) we are only able to measure \ac{ber}s down to $10^{-4}$ due to the length of the received sequence. Receiver-based noise loading on the other hand allows for the estimation of \ac{ber}s down to much lower levels as it is possible to noise load a single transmitted trace with many different noise realisations to build up the statistics. In these results we used $50000$ different noise realizations in order to measure post-\ac{fec} BERs between $10^{-7}$ and $10^{-8}$. The results for noise loading at the transmitter are in agreement with the results for noise loading at the receiver, and therefore, from now on we only consider noise loading at the receiver for post-\ac{fec} results.

Another method of comparing the performance of the two schemes, is using \acp{air}, as shown in Fig.~\ref{fig:th_info}.
Here, the thick line is the \ac{awgn} capacity $\log_2\left( 1 + \text{SNR} \right)$ \cite{Shannon1948ACommunications}.
We also consider the \ac{air} for \ac{8psk} and an AWGN channel, using the expression for \ac{mi} in  \eqref{eq:mi} and for \ac{gmi} in \eqref{eq:gmi}.
The curves are the results for an \ac{cs} \ac{awgn} channel, while the markers are obtained by calculating \eqref{eq:mi} and \eqref{eq:gmi} using transmitter noise loaded traces from the experimental setup and $N_\mathrm{D}=2$. 
Here, the majority of the noise is generated by an \ac{edfa} and has co-propagated with the signal for $1000$~km.
We find that the \ac{mi} and \ac{gmi} calculated from the experimentally obtained traces (triangles in Fig.~\ref{fig:th_info}) shows excellent agreement with the \ac{cs} \ac{awgn} model, indicating that the optical channel in the linear propagation regime is well approximated by the AWGN model. These results give upper bounds for \ac{8psk}-based transmissions of \ac{ttcm} and binary \ac{bicm} \ac{ldpc} respectively.

The post-\ac{fec} \acp{air} are also shown in Fig.~\ref{fig:th_info}, which saturate at $2$~bit/sym. These metrics are calculated using \eqref{eq:postsdmi} and \eqref{eq:posthdmi} for both schemes.
The curves are obtained by calculating \eqref{eq:postsdmi} for an \ac{cs} \ac{awgn} channel and the markers are obtained from the experimental setup.
We see that for $2$~bit/sym, there is only a $0.06$~dB SNR penalty between \ac{mi} and \ac{gmi}, however, for the actual codes that were implemented, we find that at the maximum achievable rate the \ac{ttcm} outperforms the \ac{ldpc} by $0.4$~dB. This difference in performance may be attributed to the suboptimality of the codes under consideration. With the code we implemented in this paper, the performance difference is larger than the difference between the respective bound.

Fig.~\ref{fig:th_info} also shows results for an \ac{hd} outer code (circles). The differences between the \ac{sd} and \ac{hd} bounds are only minor in the region of interest (near $2$~bit/sym), and thus, we conclude that only small penalties from choosing a \ac{hd} outer code are to be expected. At lower \ac{snr}, where the \ac{air} for \ac{sd} codes is significantly higher than the \ac{air} for \ac{hd} codes, one can imagine that a code with a lower code rate can approach the \ac{mi} and \ac{gmi} bounds far closer than the codes used in this paper do. Around $6.3$~dB \ac{snr} for the \ac{ttcm} scheme and around $6.8$~dB \ac{snr} for the \ac{ldpc} scheme, the difference in \ac{air} for the \ac{hd} with respect to the \ac{sd} codes is negligible.


\section{Conclusions}
\label{sec:conclusions}

In this paper, we experimentally compared the performance of nonbinary \ac{fec} based on turbo trellis-coded modulation and \ac{ldpc}-based binary \ac{fec} in terms of achievable information rates. 
These rates were evaluated using newly developed closed-form approximations for a correlated \ac{awgn} channel.

Unlike uncoded performance metrics, an information-theoretic analysis based on mutual information and generalized mutual information was shown to allow fair comparisons between different modulation strategies. 
{The \acp{air} can be compared for different modulation formats, including geometrically-shaped and probabilistically-shaped formats. Although in this paper all the gains were reported in terms of \ac{snr} (for a given \ac{air}), this does not always have to be the case. For example, the same methodology can be used to report gains in launch power or reach.}
This analysis, however, did not always exactly match the performance of the particular coded modulation implementations under consideration. This is because the information-theoretic analysis considers an idealized setup, e.g., infinite block lengths, unbounded decoding complexity, etc.

In this paper we only considered one modulation and code rate, however, we conjecture our conclusions to also hold for other spectral efficiencies. The study of different combinations of modulation and code rate is left for further investigation.

\appendix[Derivation of the AIR expressions]
\subsubsection{Derivation of the MI expression}\label{ape:mi_proof}
The \ac{mi} in \eqref{eq:mi_bound} can be approximated via Monte Carlo integration for any channel law $f_{\vect{Y}|\vect{X}}(\vect{Y}|\vect{X})$ using the received symbols which we denote as $\underline{\vect{y}}=[\vect{y}^{(1)},\vect{y}^{(2)},\ldots,\vect{y}^{(N_\mathrm{s})}]$.
In particular, this Monte Carlo approximation gives  \cite[eq.~(17)]{Alvarado2015ReplacingSystems}%
\begin{equation}
 I(\vect{X};\vect{Y}) \approx m + \frac1M \sum^M_{i=1} \condsumn \log_2 \frac{f_{\vect{Y}|\vect{X}}(\vect{y}^{(n)}|\vect{x}_i)}{\sum_{j=1}^{M} f_{\vect{Y}|\vect{X}}(\vect{y}^{(n)}|\vect{x}_j)} \label{eq:mi_analytical}
\end{equation}%
where $\mathcal{N}_i$ is given by \eqref{Ni.set}, and  $\underline{\vect{x}}=[\vect{x}^{(1)},\vect{x}^{(2)},\ldots,\vect{x}^{(N_\mathrm{s})}]$ are the transmitted symbols.
Then, by substituting \eqref{eq:2d_g} into \eqref{eq:mi_analytical} and using $\vect{y}^{(n)}-\vect{x}_i=\vect{z}^{(n)}$ and $\vect{y}^{(n)}-\vect{x}_j=\vect{z}^{(n)}+\vect{d}_{ij}$ for $n\in\mathcal{N}_i$, we obtain:
\begin{multline}
  I(\vect{X};\vect{Y}) \approx   m - \frac1M \sum^M_{i=1}\frac1{|\mathcal{N}_i|}  \\
  \sum_{n\in\mathcal{N}_i} \log_2 \frac{\sum_{j=1}^{M} \exp\left( -\frac12 (\vect{z}^{(n)}+\vect{d}_{ij})^\mathrm{T} \matinv{\Sigma} (\vect{z}^{(n)} +\vect{d}_{ij}) \right)}{\exp\left(-\frac12 (\vect{z}^{(n)})^\mathrm{T} \matinv{\Sigma} \vect{z}^{(n)} \right)}.\label{eq:mi_start_proof}
\end{multline}
{Rewriting the argument of the logarithm in \eqref{eq:mi_start_proof}, combining the exponents, and using the distributive property of matrix multiplications, the argument of the resulting exponential is rewritten as}
\begin{multline}
   \sum_{j=1}^{M} \exp \biggl( \frac{1}{2} \biggl(  (\vect{z}^{(n)})^\mathrm{T} \matinv{\Sigma} \vect{z}^{(n)} - (\vect{z}^{(n)})^\mathrm{T} \matinv{\Sigma} \vect{z}^{(n)} \\
   - \vect{d}_{ij}^\mathrm{T} \matinv{\Sigma} \vect{d}_{ij}-(\vect{z}^{(n)})^\mathrm{T} \matinv{\Sigma} \vect{d}_{ij} - \vect{d}_{ij}^\mathrm{T} \matinv{\Sigma} \vect{z}^{(n)} \biggr)\biggr).
\label{eq:g_z} 
\end{multline} %
Any covariance matrix is Hermitian positive-definite, and thus, $(\vect{z}^{(n)})^\mathrm{T} \matinv{\Sigma} \vect{d}_{ij} =\vect{d}_{ij}^\mathrm{T} \matinv{\Sigma} \vect{z}^{(n)}$. Using this with \eqref{eq:g_z} in \eqref{eq:mi_start_proof} gives \eqref{eq:mi}.

\subsubsection{Derivation of the GMI expression}\label{ape:gmi_proof}

The \ac{gmi} in \eqref{eq:gmi_analytical} can be approximated via Monte Carlo integration as %
\begin{multline}
 \mathrm{GMI} \approx \,    m + \frac{1}{M}\sum^m_{k=1} \sum_{l\in\lbrace0,1\rbrace} \sum_{i\in\mathcal{I}_{l,k}}   \frac1{|\mathcal{N}_i|}  \\
  \sum_{n\in\mathcal{N}_i} \log_2 \frac{\sum_{j\in\mathcal{I}_{l,k}} f_{\vect{Y}|\vect{X}}(\vect{y}^{(n)}|\vect{x}_j^{(n)}) }{\sum_{p=1}^M  f_{\vect{Y}|\vect{X}}(\vect{y}^{(n)}|\vect{x}_p^{(n)})}\label{eq:gmi_montecarlo}
\end{multline}%
where $\mathcal{N}_i$ is given by \eqref{Ni.set} and $\mathcal{I}_{l,k}$ by \eqref{Nlk.set}.
In analogy to \eqref{eq:mi_analytical}, the expression in \eqref{eq:gmi_montecarlo} is a Monte Carlo approximation of the \ac{gmi} for any channel law. The expression in \eqref{eq:gmi} is obtained by using \eqref{eq:2d_g} in \eqref{eq:gmi_montecarlo} and by following steps similar to \eqref{eq:mi_start_proof}--\eqref{eq:g_z}.

\section*{Acknowledgments}

This work was financially supported by the Engineering and Physical Sciences Research Council (EPSRC) project UNLOC (EP/J017582/1) UK, and by the Nederlandse Organisatie voor Wetenschappelijk Onderzoek (NWO) under Visitor's Travel Grant number 040.11.550/880.
Eric Sillekens is supported by the UK Engineering and Physical Sciences Research Council (EPSRC) grant EP/M507970/1 and Xtera Communications Inc. The authors would like to thank John van Weerdenburg and Dr. Roy van Uden for their support during the experiments.

\balance

\ifCLASSOPTIONcaptionsoff
  \newpage
\fi

\bibliographystyle{IEEEtran}
\bibliography{IEEEabrv,refs}
\end{document}